\documentclass[aps,prl,showpacs,twocolumn,floatfix]{revtex4}

\usepackage{amsfonts}
\usepackage{epic,eepic}
\usepackage{graphicx}

\newcommand{\tr}{\mbox{\rm Tr}}

\renewcommand{\phi}{\varphi}
\renewcommand{\rho}{\varrho}
\renewcommand{\epsilon}{\varepsilon}
\renewcommand{\theta}{\vartheta}

\begin{document}

\title {Diffusive-ballistic Crossover in 1D Quantum Walks}

\author{Daniel K. W\'{o}jcik} \email{danek@cns.physics.gatech.edu}
\homepage{http://www.cft.edu.pl/~danek}
\affiliation{Institute for Physical Science and Technology 
  University of Maryland,
  College Park, MD, 20742, USA}
\affiliation{Centrum Fizyki Teoretycznej Polskiej Akademii Nauk,
  Al. Lotnik\'{o}w 32/46, 02-668 Warszawa, Poland}
\author{J. R. Dorfman} \affiliation{Institute for Physical Science and
  Technology, and Department of Physics, University of Maryland,
  College Park, Maryland 20742}


\date{\today}

\begin{abstract}
  We show that particle transport, as characterized by the equilibrium
  mean square displacement, in a uniform, quantum multi-baker map, is
  generically ballistic in the long time limit, for any fixed value of
  Planck's constant.  However, for fixed times, the semi-classical
  limit leads to diffusion.  Random matrix theory provides explicit
  analytical predictions for the mean square displacement of a
  particle in the system. These results exhibit a crossover from
  diffusive to ballistic motion, with crossover time on the order of
  the inverse of Planck's constant. This diffusive behavior is a
  property of the equilibrium average and does not require further
  interactions of the system with the environment.  We expect that,
  for a large class of 1D quantum random walks similar to the quantum
  multi-baker, a sufficient condition for diffusion in the
  semi-classical limit is classically chaotic dynamics in each cell.
  These results describe an interesting generalization of known
  quantum random walks and may have applications for quantum
  computation.
\end{abstract}

\maketitle

Recent results for non-equilibrium and transport properties of
extended classical systems with microscopic
chaos~\cite{dorfman99s,gaspard98sa} have suggested that one might
explore the transport behavior of quantum versions of simple, extended
classical systems. One such system, studied here, is the multi-baker
model for deterministic diffusion in one-dimension.  In
paper~\cite{wojcik02da} we introduced a quantum version of the
classical multi-baker model as a combination of local quantum baker
dynamics with a transport of wave functions to neighboring cells.  Its
simplest case is an example of the quantum random walks that have been
considered previously (the Hadamard
walk)~\cite{GodoyG96,travaglione02f}, but other cases form a much
wider class of 1D quantum random walks.  Quantum random walks have
become of considerable interest over the past few years since they
might be useful for implementing quantum random search algorithms if
quantum computation becomes possible.  Quantum multi-baker models
might be excellent candidates for implementation as part of current
efforts to provide techniques and algorithms designed to take
advantage of the possibilities inherent in quantum computation. Its
experimental realization eventually may be feasible since both quantum
baker maps~\cite{BrunS99,weinstein02f}, and the Hadamard
walks~\cite{travaglione02f}, are now within experimental reach.  The
general case we consider allows a larger number of quantum states
participating in the random walk, and widens the range of possible
physical applications.

In this Letter we report on the behavior of the equilibrium mean
square displacement (m.s.d.) for a quantum particle whose classical
dynamics is governed by the deterministic, diffusive multi-baker
process.  Since one-dimensional processes in extended quantum systems
show ballistic motion, for translationally invariant systems, or
localization, for disordered ones, it is interesting to study how each
system makes the transition to diffusive behavior in the classical
limit. Here we concentrate on the translationally invariant case.  A
natural question is whether or not an interaction with the external
world, or decoherence, is necessary to eventually restore diffusive
motion in the classical limit. Our work here shows that this is not
the case for the m.s.d.: for systems that we study there is a
crossover from short-time diffusive behavior to ballistic motion in
the long time limit. The crossover time is on the order of the inverse
Planck constant, i.e. the Heisenberg time, rather than the Ehrenfest
time, on the order of the log of Planck's constant. The chaotic
classical motion of our system allows the use of random matrix theory
(RMT), which leads to explicit expressions for the mean square
displacement of a quantum particle.  The comparison of the RMT
analytic results with results of numerical evaluation of the exact
formula for particular systems is very good with some interesting
exceptions. Our calculation has a classical
counterpart~\cite{gaspard98sa} which has the same result for the
m.s.d. as obtained here in the semi-classical limit.

We begin with the classical multi-baker
map~\cite{gaspard92s,tasaki95s}.  It is a deterministic model of the
random walk on a one-dimensional lattice, where the direction of the
jump is determined by the internal state of the particle. We use the
baker map as a model of internal dynamics.  That is, the multi-baker
map, $M$, is a composition of two maps $M = B \circ T$. First, the
phase points are transported to neighboring cells according to
$T(n,x,y) = (n+1, x, y)$, for $0 \leq x < 1/2$, and $(n-1, x, y)$ for
$1/2 \leq x < 1$.  Then a baker map $B$ acts on the $x,y$ coordinates
of each cell, $n$, separately, according to $B(n,x,y) = (n, 2x, y/2)$,
for $0 \leq x < 1/2$, and $(n, 2x-1, (1+ y)/2)$, for $1/2 \leq x < 1$.
The combination of these two maps is the multi-baker map which is a
time-reversible, measure preserving, chaotic transformation, with
evolution law
\[
  M(n,x,y) =
  \left\{
    \begin{array}[c]{ll}
      (n+1, 2x, \frac{y}{2}), &\,{\rm for} \; 0 \leq x < 1/2, \\
      (n-1, 2x-1,  \frac{1+ y}{2}), &\, {\rm for}\; 1/2 \leq x < 1.
    \end{array}
  \right.
\]
It is the simplest area-preserving model of a random walk.

To quantize the dynamics we first quantize $B$ in a single cell. The
horizontal direction of the torus $[0,1)^2$ is taken to be the
coordinate axis, while vertical axis corresponds to the momentum
direction. To obtain the Hilbert
space~\cite{Hannay80,Saraceno90,DeBievreEG96} we take the subspace of
the wave functions on a line whose probability densities,
$|\Psi(x)|^2,|\tilde{\Psi}(p)|^2$ are periodic in both position and
momentum representations, respectively: $\Psi(x+1) = \exp(i2\pi
\phi_q) \Psi(x)$, $\tilde{\Psi}(p+1) = \exp(i2\pi \phi_p)
\tilde{\Psi}(p)$, where $\phi_q, \phi_p \in [0,1)$ are phases
parameterizing quantization.  The quantization of the baker map
requires the phase space volume to be an integer multiple of the
quantum of action~\cite{Hannay80,Saraceno90,DeBievreEG96}, so the
effective Planck constant must be $h=1/N$, where $N$ is the dimension
of the Hilbert space. The space and momentum representations are
connected by a discrete Fourier transform $G_N(\phi_q, \phi_p) :=
\langle p_k|q_l\rangle = {N}^{-1/2} \exp(-i 2\pi N p_k q_l)$. The
discrete positions and momenta are $q_l = (l+\phi_q)/N$, $p_k =
(k+\phi_p)/N$. The Hilbert space $H$ can be decomposed into ``left''
and ``right'' subspace $H=H_L \oplus H_R$, and ``bottom''/``top''
spaces $H=H_B \oplus H_T$, which are $N/2$ dimensional, $\Psi \in H_L$
when $\langle q_l | \Psi\rangle = 0$ for $l=N/2, \dots,N-1$, $\Psi \in
H_B$ when $\langle p_k | \Psi\rangle = 0$ for $k=N/2, \dots,N-1$.

Having constructed the Hilbert space one looks for a family of unitary
propagators parameterized by $N=1/h$ which go over into the classical
map in semi-classical limit. Technically, one requires the Egorov
condition to be satisfied, which means semi-classical commutation of
the quantum and classical evolution~\cite{DeBievreEG96}.  The unitary
operator for the quantum baker
map~\cite{BalazsV89,Saraceno90,DeBievreEG96} is given by $U_N :=
G_N^{-1} \left[
  \begin{array}[c]{cc}
    G_{N/2} & 0 \\ 0 & G_{N/2}
  \end{array}
\right]$ for even $N$.  Other examples and discussions of issues
concerning the quantization of area-preserving maps can be found e.g.
in~\cite{baecker02f}.

Since the quantum multi-baker is a model of particle with $N$ internal
states jumping over lattice of length $L$, we use product Hilbert
space $H={\mathbb C}^L \otimes {\mathbb C}^N$ to describe it.  The
dynamics is implemented in two steps. First one shifts states from
right subspace at cell $n$ ($H_R(n)$)to right states at cell $n-1$,
and states $H_L(n)$ into left states at cell $n+1$, which gives the
quantum transport operator $T$. Then on each of the cells one acts
independently with a quantum baker operator $U_N$, which corresponds
to the classical map $B$. Let us write the states $|\Psi\rangle \in H$
in the position basis $\sum_{n=0}^{L-1}\sum_{j=0}^{N-1} \Psi_{j}(n)
|n,j\rangle$. Then the only non-zero matrix elements of the quantum
multi-baker operator $M$ are of the form $\langle n+1,j | M|
n,k\rangle = \langle j | U_N | k \rangle $, for $k\in H_L(n)$, or
$\langle n-1,j | M| n,k\rangle = \langle j | U_N | k \rangle $, for
$k\in H_R(n)$. For the translationally invariant case studied here,
the m.s.d. can be expressed entirely in terms of the properties of the
baker operator, $B$, {\it cf.} Eq. (2), and we do not need to make
explicit use of the structure of $M$.

\begin{figure}[htbp]
  \centering
  \begin{tabular}[c]{c}
   \includegraphics[scale=.5]{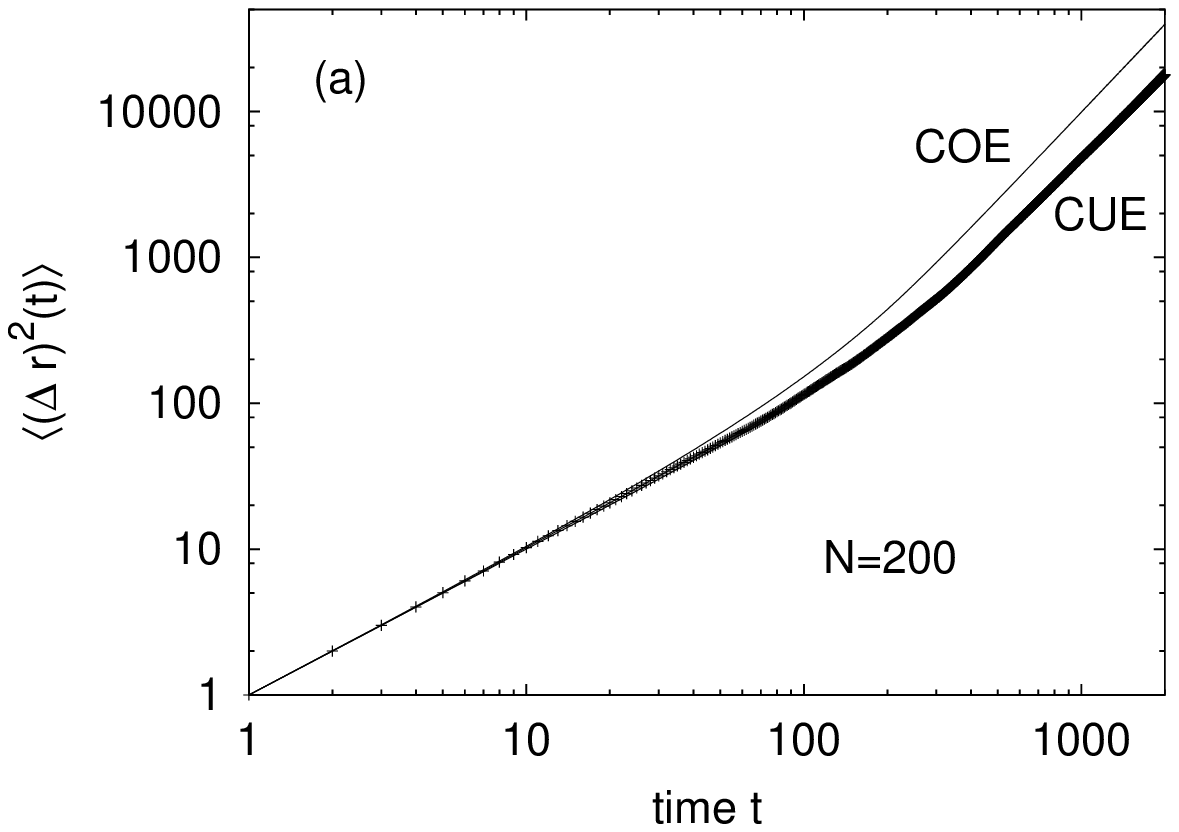} \\
   \includegraphics[scale=.5]{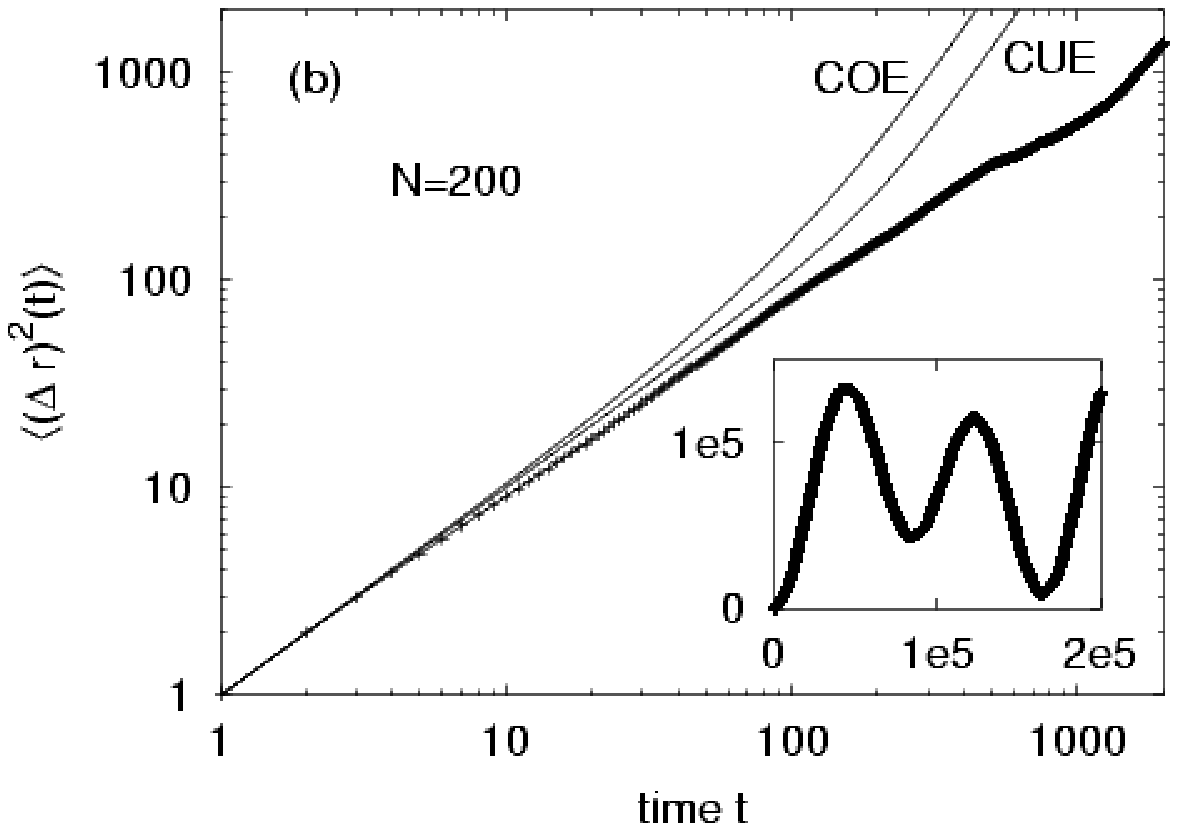}
  \end{tabular}
  \caption{
    Comparison of RMT estimates with the numerical evaluation of the
    formula~(\ref{eq:msd3}) for the m.s.d. in case of quantum
    multi-bakers with a) Balazs-Voros phases (generic case), b)
    Saraceno phases (exceptional, localized case). Plots are in double
    logarithmic scale, the inset in figure (b) shows the results for
    much longer time in normal scale.}
  \label{fig:baker}
\end{figure}

The central quantity of interest, is the expression for the mean
square displacement (m.s.d.) of a quantum particle in the chain,
defined as the average value of the mean square displacement taken
with an equilibrium density matrix for the system. This is a uniform
distribution of probabilities along the chain, $\rho_{eq}= {\mathbb
  I}_{NL}/NL$; $\langle A \rangle := \tr\ (\rho_{eq} A) =\tr\ (A)
/NL$, $L$ is the length of the chain (we assume periodic boundary
conditions and where $N$ is the dimension of the Hilbert space. Then
the m.s.d. is simply $\langle (\Delta r)^2(t)\rangle = \langle
(M^{\dagger t} r M^t -r)^2 \rangle$.  If we define the velocity
operator $v := M^\dagger r M -r$ then the m.s.d. can be written as
$\langle (\Delta r)^2(t)\rangle = \sum_{m,n=0}^{t-1} \langle v_m
v_n\rangle$, where $v_n := M^{\dagger n} v M^n$. Time invariance of
$\rho_{\rm eq}$ implies $\langle v_m v_n \rangle = \langle v_{m-n} v_0
\rangle$. Thus we can express the m.s.d. in terms of the velocity
autocorrelation function $C_n=\langle v_n v_0 \rangle$:
\begin{equation}
  \label{eq:msd}
  \langle (\Delta r)^2(t)\rangle 
   = t \langle v^2 \rangle + 2 \sum_{n=1}^{t-1} (t-n) C_n
\end{equation}
We use a ``coarse'' position operator $r$ defined by $ r |n,k \rangle
:= n |n,k\rangle $. We calculate the coarse velocity operator on the
line using the definition given above, obtaining~\cite{wojcik02dc}, $
v(n,k; m, l) = \pm \delta_{k,l} \delta_{n,m}$, with $+$ for $l<N/2$,
$-$ for other $l$, and then put it on the circle to enforce
translational invariance.  This form can be understood by observing
that for the translationally invariant case, the coarse velocity $\pm
1$ denotes a translation of the quantum state one cell to the right or
left. An identical form for the coarse velocity occurs in the
classical multi-baker as well~\cite{dorfman99s}. Then the velocity
autocorrelation function can be reduced to the trace over states in a
single cell~\cite{wojcik02dc}
\begin{equation}
  C_n
  = \frac{1}{LN}\tr\ [M^{\dagger n} v M^n v]
   = \frac1N \tr\ [B^{\dagger n} J B^n J],
\end{equation}  
where we note that $J$ is the velocity operator $v$ reduced to a
single cell.  In the position representation $J$ is given by \( J
=\left(\begin{array}{cc}
    \mathbb{I}_{N/2} & 0 \\
    0 & - \mathbb{I}_{N/2}
  \end{array}\right).
\) Assuming the properties of the local propagator, $B$, are known,
one can express the velocity autocorrelation function in terms of its
spectrum and eigenvectors, where $B |k\rangle = \exp(i \phi_k)
|k\rangle$.  It immediately follows that
\begin{equation}
  C_n = \frac{1}{N} \sum_{j,k} |J_{jk}|^2  e^{i (\phi_j-\phi_k) n}, 
 \label{cn}
\end{equation}
where $J_{jk}:=\langle j|J|k \rangle$. Since $J$ has a very simple
form, one sees that
\begin{equation}
  \sum_{j,k} |J_{jk}|^2 = \tr\ J^2 = N.
  \label{eq:sumofj}
\end{equation}
Substituting these results in formula~(\ref{eq:msd}) we obtain
\begin{eqnarray}
  \langle (\Delta r)^2(t) \rangle  
   & = & 
   \frac1N t \sum_{j\neq k} |J_{jk}|^2 +  
   \frac1N t^2 \sum_{j} |J_{jj}|^2 \nonumber \\
   && +
   \frac2N \sum_{j\neq k} |J_{jk}|^2 \sum_{n=1}^{t-1} (t-n) \,
   e^{i \alpha_{jk} n}\label{eq:msd2}\\
   & = & 
  \frac{t^2}{N} \sum_{j} |J_{jj}|^2 +
  \sum_{j\neq k}\frac{|J_{jk}|^2}{N} \frac{\sin^2 \frac{(\alpha_{jk}
  t)}{2}}{\sin^2  \frac{\alpha_{jk}}{2}}
\label{eq:msd3}
\end{eqnarray}
where $\alpha_{jk}:= \phi_j-\phi_k$. Whenever there is a degeneracy,
$\alpha_{jk}=0$, we replace the ratio of sines by $t^2$.  In certain
cases it is possible that the coefficient of the ballistic, $t^2$,
term may be zero; see Figure~\ref{fig:baker}.b and the following
discussion.

\begin{figure}[htbp]
  \centering
  \includegraphics[scale=.5]{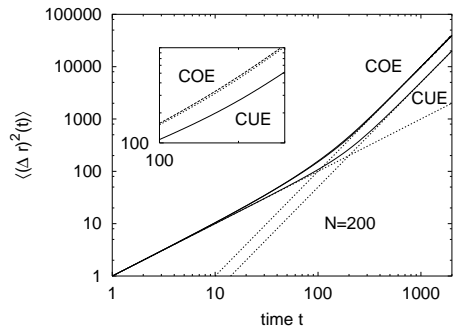}
  \caption{Log-log plot of the ensemble averages of the mean square
    displacement using RMT.  The COE results are the two close curves,
    where the lower is the result given in Eq.~(\ref{eq:cue2}) for
    $k=2$, while the higher curve was obtained using the full pair
    correlation function for COE.  Three asymptotic estimates $t,
    t^2/N, 2t^2/N$ are also plotted. The inset shows the region
    $t=100$ to $t=300$ where the differences between the two COE
    results are most pronounced. }
  \label{fig:aver}
\end{figure}

We now evaluate the m.s.d., Eq.~(\ref{eq:msd2}) using random matrix
theory and compare the results with numerical evaluations.  To apply
RMT we consider the velocity autocorrelation function $C_n$ given by
Eq. (\ref{cn}), and separate the terms on the right hand side into
diagonal, $j=k$, and non-diagonal, $j \ne k$, terms.  We suppose that
$B$ is drawn randomly from either the COE or CUE ensembles, although
the numerical results present a more general behavior.  We assume the
distribution of matrix elements is independent of the distribution of
elements of eigenvectors (see section 8.2 of~\cite{Haake01da},
and~\cite{Kus88f}).  Using $\tr J^2 = N$, one sees that the ensemble
average of the mean square displacement takes the form
\begin{eqnarray}
  \langle \langle (\Delta r)^2\rangle\rangle & = & t + t(t-1)
  \langle|J_{jj}|^2\rangle \nonumber\\ 
  && + 2 (N-1)  \langle|J_{j\neq k}|^2\rangle \sum_{n=1}^{t-1} (t-n) 
  \,\langle e^{i \alpha n} \rangle. \qquad\  
\end{eqnarray}
We replace the matrix elements $|J_{jj}|^2$ and $|J_{jk}|^2$ by their
average values $\langle|J_{jj}|^2\rangle$ and
$\langle|J_{jk}|^2\rangle$, respectively. Straightforward
calculation~\cite{wojcik02dc} gives
$\langle|J_{jj}|^2\rangle=k/(N+k)$, where $k=1$ for CUE, and 2 for
COE. Averaging Eq.~(\ref{eq:sumofj}), we obtain
$\langle|J_{jj}|^2\rangle + (N-1) \langle|J_{j\neq k}|^2\rangle = 1$
and thus $\langle|J_{j\neq k}|^2\rangle=N/[(N+k)(N-1)]$. Then we need
to calculate the average value of the exponential factor
$\exp[i(\phi_j-\phi_k)n]$. For this calculation we need the expression
for the pair correlation function $R(\phi_j,\phi_k)$ in the two
ensembles, so that we can express the average value as
\[
\langle e^{[i(\phi_i-\phi_k)n]}\rangle =\int_0^{2\pi} \!\!
\int_0^{2\pi}\!\!  d\phi_j \, d\phi_k \,e^{[i(\phi_i-\phi_k)n]}\,
\frac{R(\phi_j,\phi_k)}{N(N-1)}.
\]
These correlation functions are given in the
literature~\cite{mehta91da}. For the CUE, one finds that
\[  R(\phi_j,\phi_k) =\frac{N^2}{4 \pi^2}\left[1 -
  \frac{\sin^{2}\frac{N(\phi_j-\phi_k)}{2}}{N^2
    \sin^2\frac{(\phi_j-\phi_k)}{2}}\right].
  \label{cue1}
\]
A straightforward calculation~\cite{wojcik02dc}, leads to
\begin{equation}
\langle e^{[i(\phi_j-\phi_k)n]}\rangle  = \left\{ \begin{array}[c]{ll}  
    1 &\, {\rm for}\; n =0\\
        \frac{n-N}{N(N-1)} &\, {\rm for} \; n < N, \\ 
           0 &\, {\rm for}\; n \ge N.
         \end{array}
          \right.
\label{eq:cue2}
\end{equation}
Using this estimate for the average the result, we find that the
m.s.d. is given by
\[
\langle (\Delta r)^2(t) \rangle = \left\{ \begin{array}[c]{ll} t +
    \frac{t(t-1)}{N+k} \left[ k-1 + \frac{t-2}{3(N-1)} \right]
    &\,{\rm for}\; t\leq N, \\
    \frac{k}{N+k} t^2 + \frac{N}{3} - \frac{N(k-1)}{3(N+k)} &\, {\rm
      for}\; t > N.
 \end{array}
  \right. 
\]
Note that the ``super-ballistic'' $t^3$ term only occurs for $t\leq
N$, where it is typically less than or on the order of the linear
term, $t$. The exact result for the COE ensemble has a correction
arising from an additional term in the pair correlation function.
This correction is rather lengthy to write and is negligible for both
short and very long times, with the maximum deviation of at most five
percent occurring at $t=N$. The details will be given
elsewhere~\cite{wojcik02dc}.  Figure~\ref{fig:aver} shows the
estimates for the two ensembles for $N=200$.

We compare these predictions with numerical results.  For almost every
choice of phases $\phi_q,\phi_p$ defining the quantization, the
evaluated formula~(\ref{eq:msd3}) gives results between the COE and
CUE average predictions for $N$ greater than 100.  The Balazs-Voros
phases ($\phi_q=\phi_p=0$) \cite{BalazsV89} yield exceptionally good
agreement with CUE average for all values of $N$.
Figure~(\ref{fig:baker}.a) compares the exact result with the RMT
averages in this case.

An interesting exception to these results occurs when $\phi_q + \phi_p
=1$, e.g. for the Saraceno case ($\phi_q=\phi_p=1/2$)
\cite{Saraceno90}. For these special values of the phases, the
probability of finding the system on each half is 0.5 for every
eigenstate, so that $J_{jj}=0$, and there are no degeneracies.  Thus
for long times the m.s.d.  oscillates around the time average value,
$\sum_{j\neq k} |J_{jk}|^2/ (2N \sin^2 \frac{\alpha_{jk}}{2})$, after
an initial diffusive growth.  To see how such oscillations are
possible, consider a more general class of systems where the local
quantum baker map is replaced by an arbitrary unitary operator
$B$~\cite{wojcik02dc}. For this larger class of systems one easily
proves that $0 \leq \langle (\Delta r)^2\rangle \leq t^2$, both
classically and quantum-mechanically, and it is easy to construct
examples realizing both extrema. Moreover, if $B$ is a right-left
exchange operator, which merely replaces left and right states, the
m.s.d. oscillates between 0 and 1.  Similar phenomena over much longer
time-scales may take place in the case discussed above.
Figure~(\ref{fig:baker}.b) shows reasonable agreement with the
short-time classical diffusive behavior also in the case of Saraceno
quantization.  We expect that for large $N$ the phases should not
matter for times up to the Heisenberg time.

{\em Summary:} We have considered the mean square displacement of a
particle whose dynamics is governed by the propagator for a quantum
multi-baker map.  After deriving a general formula for the m.s.d., we
evaluated it by means of RMT and also by numerical methods. Random
matrix theory provides explicit, analytic expressions for the m.s.d.,
as well as for the velocity autocorrelation functions that determine
it.  Our study shows that these analytic expressions depend somewhat
on the nature of the circular ensembles used, either unitary or
orthogonal.  Comparison with numerical results shows that neither of
the ensembles is superior: depending on quantization phases one or the
other gives a better representation of the data, but usually the
experimental curve lies between the two predictions with numerical
ballistic coefficient having values between $1/N$ and $2/N$ for
generic (ballistic) systems.  The analytic expressions for the m.s.d.
allow us to study the transition to classical behavior in detail. We
find that, on the average, there is a smooth transition from quantum
to classical behavior as Planck's constant approaches zero, and that
for non-zero values of $h$ the classical behavior persists up to times
on the order of $h^{-1}$, and we see no need for further interactions
with the environment, for a well behaved classical limit for the
models we study.  Thus the analytic expressions for the velocity
autocorrelation functions and m.s.d.  provide a powerful tool for
studying the quantum-classical transition for simple extended systems.
Given that it is possible to realize a quantum baker map
experimentally, the quantum multi-baker map eventually may have
applications for quantum computations.

We show elsewhere that most of the results presented here are valid
for a much wider class of quantum systems~\cite{wojcik02dc}.

The authors would like to thank S. Fishman, P.  Gaspard, M. Ku\'s, J.
P. Paz, S.  Tasaki, J.  Vollmer, and K.  \.Zyczkowski for
helpful comments, and the National Science Foundation for support
under grant PHY-98-20824.


\begin{thebibliography}{22}
 \expandafter\ifx\csname natexlab\endcsname\relax\def\natexlab#1{#1}\fi
 \expandafter\ifx\csname bibnamefont\endcsname\relax
   \def\bibnamefont#1{#1}\fi
 \expandafter\ifx\csname bibfnamefont\endcsname\relax
   \def\bibfnamefont#1{#1}\fi
 \expandafter\ifx\csname citenamefont\endcsname\relax
   \def\citenamefont#1{#1}\fi
 \expandafter\ifx\csname url\endcsname\relax
   \def\url#1{\texttt{#1}}\fi
 \expandafter\ifx\csname urlprefix\endcsname\relax\def\urlprefix{URL }\fi
 \providecommand{\bibinfo}[2]{#2}
 \providecommand{\eprint}[2][]{\url{#2}}

 \bibitem[{\citenamefont{Dorfman}(1999)}]{dorfman99s}
 \bibinfo{author}{\bibfnamefont{J.~R.} \bibnamefont{Dorfman}},
   \emph{\bibinfo{title}{An introduction to chaos in nonequilibrium statistical
   mechanics}} (\bibinfo{publisher}{Cambridge University Press},
   \bibinfo{address}{Cambridge}, \bibinfo{year}{1999}).

 \bibitem[{\citenamefont{Gaspard}(1998)}]{gaspard98sa}
 \bibinfo{author}{\bibfnamefont{P.}~\bibnamefont{Gaspard}},
   \emph{\bibinfo{title}{Chaos, scattering and statistical mechanics}}
   (\bibinfo{publisher}{Cambridge University Press},
   \bibinfo{address}{Cambridge}, \bibinfo{year}{1998}).

 \bibitem[{\citenamefont{W{\'o}jcik and
   Dorfman}(2002{\natexlab{a}})}]{wojcik02da}
 \bibinfo{author}{\bibfnamefont{D.~K.} \bibnamefont{W{\'o}jcik}}
   \bibnamefont{and} \bibinfo{author}{\bibfnamefont{J.~R.}
   \bibnamefont{Dorfman}}, \bibinfo{journal}{Phys. Rev. E}
   \textbf{\bibinfo{volume}{66}}, \bibinfo{pages}{036110}
   (\bibinfo{year}{2002}{\natexlab{a}}).


 \bibitem[{\citenamefont{Godoy and Garcia-Colin}(1996)}]{GodoyG96}
 \bibinfo{author}{\bibfnamefont{S.}~\bibnamefont{Godoy}} \bibnamefont{and}
   \bibinfo{author}{\bibfnamefont{L.~S.} \bibnamefont{Garcia-Colin}},
   \bibinfo{journal}{Phys. Rev. E} \textbf{\bibinfo{volume}{53}},
   \bibinfo{pages}{5779} (\bibinfo{year}{1996});
 \bibinfo{author}{\bibfnamefont{A.}~\bibnamefont{Ambainis}},
   \bibinfo{author}{\bibfnamefont{et}~\bibnamefont{al.}},
   in \emph{\bibinfo{booktitle}{Proceedings of the 30th STOC}}
   (\bibinfo{publisher}{ACM}, \bibinfo{year}{2001}),
   pp. \bibinfo{pages}{37--49}. 


 \bibitem[{\citenamefont{Travaglione and Milburn}(2002)}]{travaglione02f}
 \bibinfo{author}{\bibfnamefont{B.~C.} \bibnamefont{Travaglione}}
   \bibnamefont{and} \bibinfo{author}{\bibfnamefont{G.~J.}
   \bibnamefont{Milburn}}, \bibinfo{journal}{Phys. Rev. A}
   \textbf{\bibinfo{volume}{65}}, \bibinfo{pages}{032310}
   (\bibinfo{year}{2002});
 \bibinfo{author}{\bibfnamefont{W.}~\bibnamefont{D\"ur}},
   \bibinfo{author}{\bibfnamefont{R.}~\bibnamefont{Raussendorfer}},
   \bibinfo{author}{\bibfnamefont{V.~M.} \bibnamefont{Kendon}},
   \bibnamefont{and} \bibinfo{author}{\bibfnamefont{H.-J.}
   \bibnamefont{Briegel}} (\bibinfo{year}{2002}),
   \bibinfo{note}{quant-ph/0207137}.

 \bibitem[{\citenamefont{Brun and Schack}(1999)}]{BrunS99}
 \bibinfo{author}{\bibfnamefont{T.~A.} \bibnamefont{Brun}} \bibnamefont{and}
   \bibinfo{author}{\bibfnamefont{R.}~\bibnamefont{Schack}},
   \bibinfo{journal}{Phys. Rev. A} \textbf{\bibinfo{volume}{59}},
   \bibinfo{pages}{2649} (\bibinfo{year}{1999}).

 \bibitem[{\citenamefont{Weinstein et~al.}(2002)\citenamefont{Weinstein, Lloyd,
   Emerson, and Cory}}]{weinstein02f}
 \bibinfo{author}{\bibfnamefont{Y.~S.} \bibnamefont{Weinstein}},
   \bibinfo{author}{\bibfnamefont{S.}~\bibnamefont{Lloyd}},
   \bibinfo{author}{\bibfnamefont{J.}~\bibnamefont{Emerson}}, \bibnamefont{and}
   \bibinfo{author}{\bibfnamefont{D.~G.} \bibnamefont{Cory}},
   \bibinfo{journal}{Phys. Rev. Lett.} \textbf{\bibinfo{volume}{89}},
   \bibinfo{pages}{157902} (\bibinfo{year}{2002}).

 \bibitem[{\citenamefont{Gaspard}(1992)}]{gaspard92s}
 \bibinfo{author}{\bibfnamefont{P.}~\bibnamefont{Gaspard}}, \bibinfo{journal}{J.
   Stat. Phys.} \textbf{\bibinfo{volume}{68}}, \bibinfo{pages}{673}
   (\bibinfo{year}{1992}).

 \bibitem[{\citenamefont{Tasaki and Gaspard}(1995)}]{tasaki95s}
 \bibinfo{author}{\bibfnamefont{S.}~\bibnamefont{Tasaki}} \bibnamefont{and}
   \bibinfo{author}{\bibfnamefont{P.}~\bibnamefont{Gaspard}},
   \bibinfo{journal}{J. Stat. Phys.} \textbf{\bibinfo{volume}{81}},
   \bibinfo{pages}{935} (\bibinfo{year}{1995}).

 \bibitem[{\citenamefont{Hannay and Berry}(1980)}]{Hannay80}
 \bibinfo{author}{\bibfnamefont{J.~H.} \bibnamefont{Hannay}} \bibnamefont{and}
   \bibinfo{author}{\bibfnamefont{M.~V.} \bibnamefont{Berry}},
   \bibinfo{journal}{Physica D} \textbf{\bibinfo{volume}{1}},
   \bibinfo{pages}{267} (\bibinfo{year}{1980}).

 \bibitem[{\citenamefont{Saraceno}(1990)}]{Saraceno90}
 \bibinfo{author}{\bibfnamefont{M.}~\bibnamefont{Saraceno}},
   \bibinfo{journal}{Ann. Phys.} \textbf{\bibinfo{volume}{199}},
   \bibinfo{pages}{37} (\bibinfo{year}{1990}).

 \bibitem[{\citenamefont{DeBievre et~al.}(1996)\citenamefont{DeBievre, Esposti,
   and Giachetti}}]{DeBievreEG96}
 \bibinfo{author}{\bibfnamefont{S.}~\bibnamefont{DeBievre}},
   \bibinfo{author}{\bibfnamefont{M.~D.} \bibnamefont{Esposti}},
   \bibnamefont{and}
   \bibinfo{author}{\bibfnamefont{R.}~\bibnamefont{Giachetti}},
   \bibinfo{journal}{Commun. Math. Phys.} \textbf{\bibinfo{volume}{176}},
   \bibinfo{pages}{73} (\bibinfo{year}{1996}).

 \bibitem[{\citenamefont{Balazs and Voros}(1989)}]{BalazsV89}
 \bibinfo{author}{\bibfnamefont{N.~L.} \bibnamefont{Balazs}} \bibnamefont{and}
   \bibinfo{author}{\bibfnamefont{A.}~\bibnamefont{Voros}},
   \bibinfo{journal}{Ann. Phys.} \textbf{\bibinfo{volume}{190}},
   \bibinfo{pages}{1} (\bibinfo{year}{1989}).

 \bibitem[{\citenamefont{B\"acker}(2002)}]{baecker02f}
 \bibinfo{author}{\bibfnamefont{A.}~\bibnamefont{B\"acker}}
   (\bibinfo{year}{2002}), \bibinfo{note}{nlin.CD/0204061}.

 \bibitem[{\citenamefont{W{\'o}jcik and
   Dorfman}(2002{\natexlab{b}})}]{wojcik02dc}
 \bibinfo{author}{\bibfnamefont{D.~K.} \bibnamefont{W{\'o}jcik}}
   \bibnamefont{and} \bibinfo{author}{\bibfnamefont{J.~R.}
   \bibnamefont{Dorfman}} (\bibinfo{year}{2002}{\natexlab{b}}),
   \bibinfo{note}{in preparation}, and \bibinfo{note}{nlin.CD/0212036}.

 \bibitem[{\citenamefont{Haake}(2001)}]{Haake01da}
 \bibinfo{author}{\bibfnamefont{F.}~\bibnamefont{Haake}},
   \emph{\bibinfo{title}{Quantum Signatures of Chaos}}
   (\bibinfo{publisher}{Springer Verlag}, \bibinfo{address}{Berlin, New York},
   \bibinfo{year}{2001}), \bibinfo{edition}{2nd} ed.

 \bibitem[{\citenamefont{Ku\'s et~al.}(1988)\citenamefont{Ku\'s, Mostowski, and
   Haake}}]{Kus88f}
 \bibinfo{author}{\bibfnamefont{M.}~\bibnamefont{Ku\'s et al.}},
   \bibinfo{journal}{J. Phys. A} \textbf{\bibinfo{volume}{21}},
   \bibinfo{pages}{L1073} (\bibinfo{year}{1988}).

\bibitem[{\citenamefont{Mehta}(1991)}]{mehta91da}
\bibinfo{author}{\bibfnamefont{M.~L.} \bibnamefont{Mehta}},
  \emph{\bibinfo{title}{Random matrices}} (\bibinfo{publisher}{Academic Press},
  \bibinfo{address}{New York}, \bibinfo{year}{1991}).

\end{thebibliography}
\end{document}